\documentclass[conference]{IEEEtran}
\IEEEoverridecommandlockouts
\usepackage{amsthm}
\usepackage{setspace}
\usepackage{xcolor}
\newtheorem{theorem}{Theorem}

\newtheorem{lemma}{Lemma}
\newtheorem{definition}{Definition}

\newtheorem{assumption}{Assumption}
\usepackage{cite}
\usepackage{amsmath,amssymb,amsfonts}
\usepackage{algorithmic}
\usepackage{graphicx}
\usepackage{textcomp}
\usepackage{xcolor}
\usepackage{listings}
\usepackage{hyperref}
\usepackage{tikz}
\usepackage{main}
\usepackage{cleveref}
\usepackage{breqn}
\usepackage{flushend}
\usetikzlibrary{positioning,chains,fit,shapes,calc}

\graphicspath{{./graphics/}}

\def\BibTeX{{\rm B\kern-.05em{\sc i\kern-.025em b}\kern-.08em
    T\kern-.1667em\lower.7ex\hbox{E}\kern-.125emX}}

    \title{Coded Kalman Filtering Over Gaussian Channels with Feedback\\
}

\begin{document}

\author{\IEEEauthorblockN{Barron Han$^1$, Oron Sabag$^2$, Victoria Kostina$^1$, Babak Hassibi$^1$}
\IEEEauthorblockA{\textit{$^1$California Institute of Technology, $^2$The Hebrew University of Jerusalem}\\
bshan@caltech.edu}
}
\maketitle

\begin{abstract}
This paper investigates the problem of zero-delay joint source-channel coding of a vector Gauss-Markov source over a multiple-input multiple-output (MIMO) additive white Gaussian noise (AWGN) channel with feedback. In contrast to the classical problem of causal estimation using noisy observations, we examine a system where the source can be encoded before transmission. An encoder, equipped with feedback of past channel outputs, observes the source state and encodes the information in a causal manner as inputs to the channel while adhering to a power constraint. The objective of the code is to estimate the source state with minimum mean square error at the infinite horizon. This work shows a fundamental theorem for two scenarios: for the transmission of an unstable vector Gauss-Markov source over either a multiple-input single-output (MISO) or a single-input multiple-output (SIMO) AWGN channel, finite estimation error is achievable if and only if the sum of logs of the unstable eigenvalues of the state gain matrix is less than the Shannon channel capacity. We prove these results by showing an optimal linear innovations encoder that can be applied to sources and channels of any dimension and analyzing it together with the corresponding Kalman filter decoder.
\end{abstract}

\section{Introduction} \label{intro1}

\begin{figure*}
\begin{center}
    \includegraphics[width=\textwidth]{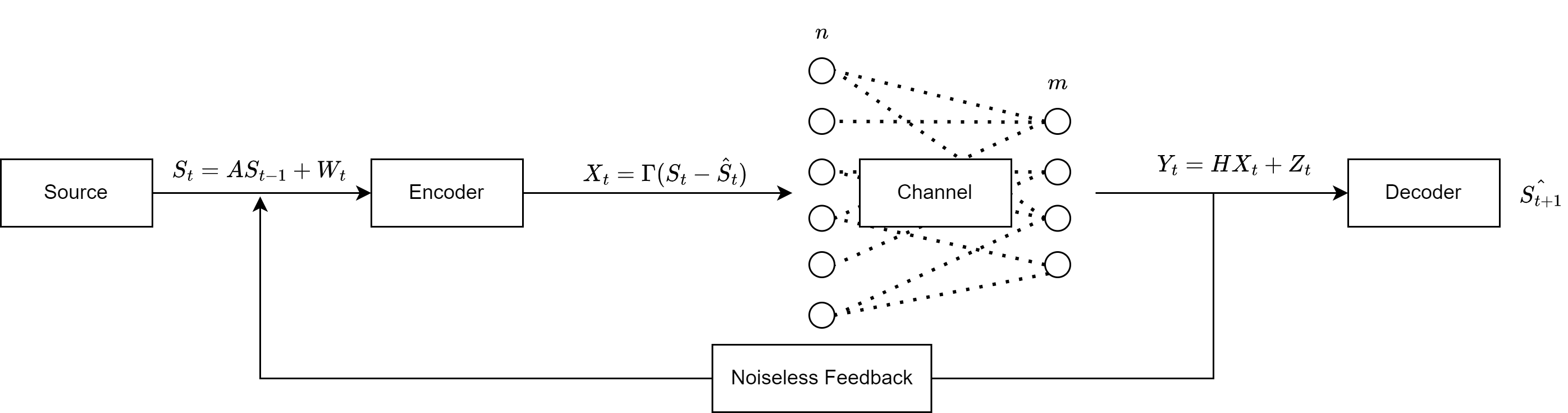}
\end{center}
\caption{A MIMO AWGN channel, described in Definition \ref{channel}, with a noiseless feedback link is shown. The Gauss-Markov source, described in Definition \ref{source}, produces information at every time, which is encoded and passed through the channel. The decoder seeks to estimate the source at the next time given all channel outputs. We show in Lemma \ref{optimalencoder} that the encoding structure displayed here is optimal for our performance metric.}
\label{fig: mimo}
\end{figure*}

To support emerging 6G communications, Internet of Things devices, and autonomous vehicles, wireless communication systems need to handle large amounts of data produced in a streaming fashion with minimal delay. For instance, unmanned drones require precise position control, which relies on potentially noisy bilateral communication between the drone and the controller. In this communication system, the source symbols are made available gradually in a streaming fashion, and the encoder has access to channel feedback. The penalty is zero-delay estimation error. Codes that are suitable to this setting do not operate on blocks. Rather, the encoder and decoder apply an immediate, recursive operation to every emergent source symbol and channel output.

The traditional communication system, introduced by Shannon, encodes data into blocks before transmitting them through a noisy channel \cite{shannon}. Since the performance criterion is asymptotic, the underlying assumption is that a long block of data symbols to be transmitted is available before encoding. These classical sources can be transmitted using block codes with a fixed block length such as \cite{hamming, reed, reedmuller, polar, ldpc}, or variable block length codes such as \cite{horstein, SK2, NaghshvarSED}, which use feedback to determine future channel inputs and adaptive decide the decoding time depending on the channel outputs. A feedback code operates on a pair of channels: the forward channel carries signals from the encoder to the decoder and is often noisy; the reverse channel carries signals from the decoder to the encoder and is assumed to be noiseless \cite{shannonfeedback}. This assumption is reasonable if the decoder has access to more power than the encoder relative to the noise floor. The well-known Schalkwijk and Kailath (SK) scheme \cite{SK2} over AWGN channels transmits a fixed message, mapped to the reals, on the first channel use, and uses subsequent channel uses to refine the decoder's estimate of the noise added during the first transmission so that it can recover the initial transmission virtually noiselessly.

The schemes mentioned so far operate on a known and fixed message. If the message is made available gradually, block encoders inevitably introduce buffering, resulting in inferior performance compared to recursive codes \cite{nian2}. Generalization of the SED code for streaming sources have been studied in \cite{nian}, while \cite{sahaimitter, Khina, elia} study a generalization of the SK scheme but only for scalar sources and channels. Floor et al. design a joint source-channel code for two correlated Gaussian processes across a multiple access channel, but they assume that the source process is memoryless \cite{skoglund}. In joint source-channel coding, the two tasks of compressing the source and encoding it for the channel are done jointly instead of with separate source and channel coders.

A special class of tree codes such as those studied by Schulman \cite{schulman} for coding evolving sources over discrete memoryless channels are recursive but do not utilize feedback. Sukhavasi et al. \cite{sukhanytime} show sufficient conditions on the rate and reliability of tree codes to estimate an unstable linear system using a random linear tree code ensemble.

This paper concerns causal transmission of a source across a particular communication channel. Conditions for accurately estimating the source involve both the source and channel characteristics. In the classical block coding setting, the relevant source and channel characteristics are the source rate-distortion function and the channel capacity, according to Shannon’s source-channel separation theorem \cite{shannon}. In the zero-delay causal setting this paper is concerned with, the corresponding fundamental source and channel characteristics are a topic of ongoing research. The source generates new bits at every time and the channel must be able to reliably carry these new bits to the encoder with a low-enough delay. Intuitively, if the source produces more information than the channel can reliably carry, it becomes impossible to accurately estimate the source. In their seminal paper, \cite{sahaimitter} argued that a more stringent requirement than classical Shannon capacity is necessary to properly describe the relationship between communication rate and error for these systems. They derived sufficient conditions to stabilize an unstable source across a communication channel in terms of anytime capacity and provide a converse result for codes over certain discrete memoryless channels that perform source quantization and channel coding separately. The causal rate-distortion function \cite{gorbunov} provides a lower bound to the channel capacity necessary for casually estimating the source subject to a given distortion over this channel \cite{kostinahassibirate}. A classical work by Gorbunov and Pinsker \cite{gorbunov} derived the sequential rate-distortion of a scalar Gauss-Markov source. Tanaka et al. \cite{Tanaka} expressed the causal rate distortion function of a vector Gauss-Markov source as a semidefinite program. Tatikonda et al. They design the parameters and dimension of an optimal linear observer of the source and analyze the directed mutual information and achievable distortion \cite{Tanaka}. This is different from a system with a given MIMO channel, which fixes the dimension of the observation. \cite{tatikonda} demonstrated the relevance of a causal rate distortion function to the problem of controlling a linear stochastic system over a noisy channel. In that setup, the encoder intelligently forms channel inputs to optimize the control objective; the decoder observes a noisy channel output and forms the control action. The lower bound to channel capacity provided by the causal rate-distortion function is known to be tight only in the rare event of a source coincidentally and fortuitously matched \cite{gastpar} to the channel at hand. For example, a scalar Gauss-Markov source is matched to the scalar AWGN channel \cite{gorbunov, sahaimitter}. In this source-channel matched setting, the optimal encoder transmits the rescaled estimation error at each time with the innovation of the source \cite{tatikonda}. In the ubiquitous scenario of the source not being matched to the channel, more sophisticated coding schemes are in order and they are a topic of current research \cite{Khina, kostinahassibirate, nian, nian2, sahaimitter}.

 Other authors analyzed the problem of controlling unstable plants under rate and cost constraints \cite{nair, nair2, borkar} or over noisy communication channels \cite{Khina, basar, franceschetti, gupta, zaidi2013stabilization}. The control problem is closely related to the estimation problem since the controller must accurately estimate the unstable source to generate control inputs. \cite{Khina} consider a single-input single-output (SISO) plant and derived conditions for a joint-source channel code to almost surely stabilize the source, while \cite{zaidi2013stabilization} consider the same setting with a vector source.
 
This paper studies zero-delay joint source-channel coding problem of a vector Gauss-Markov source over an AWGN channel. Consider a time-sensitive remote estimation system comprising a source, an encoder, a communication channel with feedback, and a decoder as in Figure \ref{fig: mimo}. The encoder leverages available source realizations and feedback of past channel outputs to generate channel inputs, while the decoder estimates the sequence of source realizations using only the channel outputs. The cost has zero delay; it measures the decoder's estimation error at every time.

We propose that a time-invariant linear encoder requires the minimum capacity to achieve finite asymptotic estimation error when transmitting a vector-valued source over a rank-one channel. We further refine the structure of the encoder to a simplified form where the channel input is a linear transformation of the recent state estimation error. This simplified structure induces an optimal Kalman filter decoder, whose performance is analyzed using classical theories in linear estimation. This code is practical since the encoder and decoder must only compute the mean square error using the Kalman filter at every time. The optimal encoder is determined by a one-time calculation of the best time-invariant matrix pre-factor which is used to encode the estimation error at every time. At the infinite horizon, we argue that the estimation error is the solution of a discrete algebraic Riccati equation (DARE) under a power constraint, which holds in the general case where the source and channel have arbitrary dimension. In the special cases of a vector source with a single-input multiple-output (SIMO) channel or a input single-output (MISO) channel, the Riccati equation reduces to a Lyapunov equation which we can analyze explicitly to derive necessary and sufficient conditions for the estimation error to be finite. We conclude that the relevant measure of source instability is the sum of the logarithms of the unstable eigenvalues of the state gain matrix. The operational channel capacity in this setting is the Shannon capacity for both SIMO and MISO channels.

% we then derive necessary and sufficient conditions for the source estimation error to be finite at the infinite horizon and connect the eigenvalues of the source gain matrix to the channel capacity of  in Theorem \ref{thm: SIMO} and Theorem \ref{thm: MISO}.

The remainder of the paper is organized as follows. Section \ref{intro} specifies the source and channel models and formally defines the class of zero-delay joint source-channel codes and the performance criterion. Section \ref{mainresult} presents our main contributions on the necessary and sufficient conditions for finite estimation error of vector source with MISO and SIMO channels. Section \ref{proofs} shows the structure of the optimal code and contains proof sketches of the main results.

We denote by $\{\bfX_t\}_{t=0}^T $ a discrete time random process and we denote the vector $X^t  \triangleq \{x_0, x_1, \ldots, x_t\}$. We write $\bfX \sim \mathcal N (\mu, \Sigma)$ to say that the random vector $\bfX$ has a Gaussian distribution with mean $\E[\bfX] = \mu$ and covariance matrix $\Cov[\bfX] = \Sigma$. Matrices and vectors are denoted with uppercase letters while scalars are denoted with lower case mathematical font. Sets are denoted using calligraphic font.

\section{Problem Setup} \label{intro}
The channel in Figure \ref{fig: mimo} is a multiple-input multiple-output (MIMO) AWGN channel.

\begin{definition}\label{channel}
(MIMO AWGN Channel) The channel accepts a vector input $X_t \in \mathbb{R}^n$ and produces a vector output $Y_t \in \mathbb{R}^m$,
\begin{equation}
    \bfY_t = H \bfX_t + \bfZ_t, \ t \ge 1
\end{equation}
where $ H \in \mathbb{R}^{m \times n}$ the channel gain matrix, $\bfZ_t \sim \mathcal{N} (0, R).$
\end{definition}

The streaming source in Figure \ref{fig: mimo} is a Gauss-Markov source $S_t$.
\begin{definition}\label{source}
(Gauss-Markov Source) The Gauss-Markov source evolves according to the linear dynamical system equation:
\begin{equation} \label{stateupdate}
    \bfS_{t+1} = A \bfS_t + \bfW_t, 
\end{equation}
where $A \in \mathbb{R}^{k \times k}$, $\bfW_t \sim \mathcal{N}(0,Q)$ and the initial state is $\bfS_0 \sim \mathcal{N} (0, Q)$.
\end{definition}

Without loss of generality, let us consider that 
\begin{equation} \label{jordanA}
    A = \left[ \begin{array}{cc} A_s & 0 \\ 0 & A_u \end{array} \right]
\end{equation}
in Jordan form in \eqref{stateupdate}, where $A_s$ is stable and $A_u$ is unstable.

We make the following assumptions about our system:
\begin{assumption} \label{asm: controllable}
The pair $(A,Q)$ is controllable.
\end{assumption}
\begin{assumption} \label{asm: eigs}
The eigenvalues of $A$ satisfy $\lambda_i \lambda_j \neq 1, \forall i,j.$
\end{assumption}
\begin{assumption} \label{asm: time_inv_enc}
We assume the encoder in Definition \ref{code} is time-invariant, meaning $\Gamma_t = \Gamma, \forall t$ in \eqref{optencoding}.
\end{assumption}
\begin{assumption} \label{asm: diag}
 We will assume that $A_u$ is diagonal in \eqref{jordanA}.
\end{assumption}

 Assumption \ref{asm: controllable} guarantees that the error covariance is positive definite. Assumption \ref{asm: eigs} guarantees that a Lyapunov equation of the form $X = AXA^T + W$ has a unique solution \cite[Lemma D.1.1]{linearestimation}. These assumptions are common in classical linear estimation theory \cite[App. C]{linearestimation} and will be utilized throughout the proofs in Section \ref{proofs}. Assumption \ref{asm: time_inv_enc} simplifies the expression for the infinite horizon estimation error, but we claim that a time-varying encoder cannot do better for our setting. Assumption \ref{asm: diag} yields a cleaner analysis for the proof of Theorem \ref{thm: MISO}, but our results hold in the general case where $A_u$ is in Jordan block form.
 
% We use the channel and source in Definition \ref{channel} and \ref{source} and define the prior minimum mean square estimates of $S$ as
% \begin{equation}
%     \hat{S}_{t} \triangleq \E[S_t|Y^{t-1}]
% \end{equation}
% with error covariance
% \begin{equation}
%     P_{t} \triangleq \Cov[S_t - \hat S_t |Y^{t-1}]. \\ 
% \end{equation}
% \textcolor{red}{This part can come after the codebook definition}
\begin{definition}\label{code}
(A Zero-Delay Joint-Source Channel Feedback Code) The feedback code for the source-channel pair in Definitions \ref{channel} and \ref{source} consists of the following:
\begin{enumerate}
  \item An encoder that at time $t$ has access to  $S^t$ and $Y^{t-1}$ and generates
  \begin{equation}
      \bfX_t = f_t(\bfS^t, \bfY^{t-1}),
  \end{equation}
  where
\begin{equation} \label{enc_def}
    f_t: \mathcal{S}^t  \times \mathcal{Y}^{t-1} \mapsto \mathbb{R}^n,
\end{equation}
    and $\mathcal S = \mathbb{R}^k, \mathcal{Y} = \mathbb{R}^m$.
    The channel inputs must satisfy a long-term average power constraint,
    \begin{equation} \label{powerconstraint}
       \frac{1}{T} \sum_{t=0}^{T-1} \E[\bfX_t^T \bfX_t] \leq p,
    \end{equation}
    up to a horizon $T$.
  \item A decoder that at time $t$ predicts the next source state,
  \begin{equation} \label{dec_def}
      \hat S_{t+1} = g_t(\bfY^t)
  \end{equation}
\end{enumerate}
\end{definition}

For a given code in Definition \ref{code}, we denote the predicted error covariance
\begin{equation} \label{cov}
    P_t \triangleq \Cov (\bfS_t - \hat \bfS_t).
\end{equation}
Note that for an encoder $f \triangleq \{f_t\}_{t=0}^\infty$ \eqref{enc_def}, the decoder
\begin{equation} \label{decoder}
    \hat\bfS_t = \E[\bfS_{t+1} | \bfY^t]
\end{equation}
minimizes the mean-square error 
\begin{equation}
    D_t \triangleq \Tr \left(\Cov(\bfS_t - \hat \bfS_t) \right).
\end{equation}
Denote the asymptotic mean square error
\begin{equation} \label{distortion}
    D \triangleq \text{limsup}_{t \to \infty} D_t. 
\end{equation}
In this paper, we examine conditions for there to exist an encoder and decoder pair such that $D < \infty$.

\section{Main Results} \label{mainresult}
We present the fundamental relation between finite estimation errors and the channel capacity for two scenarios. Theorem \ref{thm: MISO} deals with a vector source and a MISO channel. In this scenario, the channel gain, $H \in \mathbb{R}^{1 \times n}$, is normalized as $\|H\|=1$ without loss of generality.

\begin{theorem} \label{thm: MISO}
(MISO Channel) In zero-delay JSCC (Def. \ref{code}) of a vector Gauss-Markov source (Def. \ref{source}) and MISO AWGN channel, i.e., $H \in \mathbb{R}^{1 \times n}$ with $|| H || = 1$ (Def. \ref{channel}), finite asymptotic error, $D < \infty$ \eqref{distortion}, is achievable if and only if 
\begin{equation} \label{miso_res}
    \sum_{i=1}^k \log \left( \max (1, |\lambda_i|) \right) < C
\end{equation}
where $\{\lambda_i\}$ are the eigenvalues values of $A$ and $$C = \frac{1}{2} \log \left(1 + \frac{p}{r}\right)$$ is the Shannon capacity of the channel. 
\end{theorem}

\begin{figure}
\begin{center}
    \includegraphics[width=0.48\textwidth]{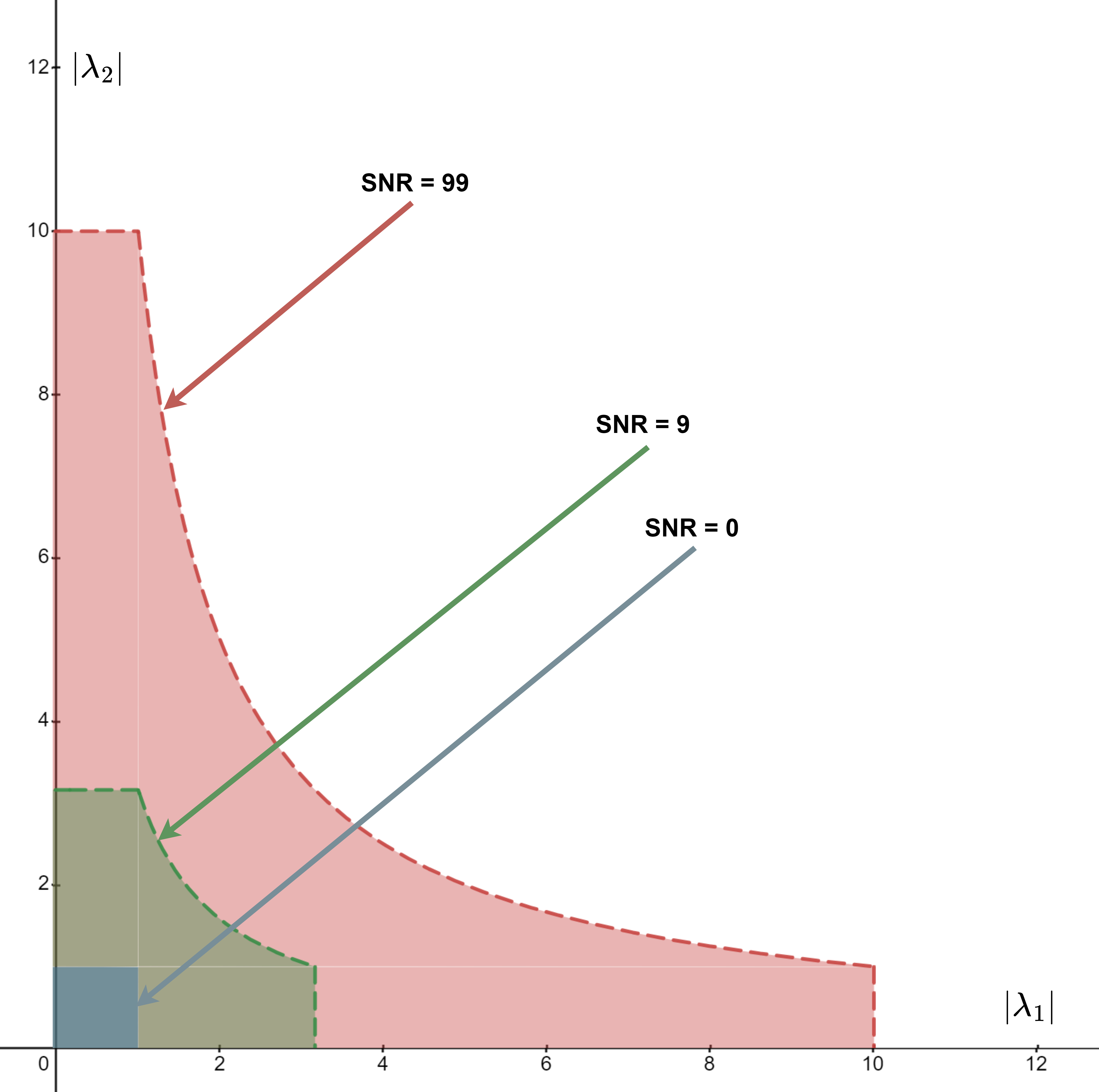}
\end{center}
\caption{Consider transmission of a 2-dimensional Gauss-Markov source over a MISO channel. The axes correspond to the eigenvalues of $A$ and the enclosed regions display where a finite estimation error is achievable for SNR $\triangleq \frac{p}{r} = 0, 9, 99$. The blue region corresponds to 0 capacity, i.e. SNR $= 0$, the decoder can only estimate a stable source.}
\label{fig: region}
\end{figure}

In the setting of Theorem \ref{thm: MISO}, the decoder estimates all $k$ components of the source using a single channel output. The region where finite estimation error is achievable is displayed in Figure \ref{fig: region} for a 2-dimensional source.

This theorem is reminiscent of a result by Elia \cite[Th. 4.3]{elia}, which involves the unstable eigenvalues of the system as well, but their result holds only for single-output multiple-input sources with different dynamics compared to our Definition \ref{source}. The adjacent control problem has also been studied by \cite{zaidi2013stabilization}, which shows that the same capacity condition is sufficient for stabilizing an unstable plant.

Next, we present the other scenario for a vector source and SIMO channel.

\begin{theorem} \label{thm: SIMO}
(SIMO) In zero-delay JSCC (Definition \ref{code}) of a vector Gauss-Markov source (Definition \ref{source}) over a SIMO ($H \in \mathbb{R}^{m \times 1}$) AWGN channel (Definition \ref{channel}), finite asymptotic error, $D < \infty$ \eqref{distortion}, is achievable if and only if 
\begin{equation}
    \sum_{i=1}^k \log \left( \max (1, |\lambda_i|) \right) < C,
\label{simo_res}
\end{equation}
where $\{\lambda_i\}$ are the eigenvalues values of $A$ and $$C = \frac{1}{2} \log \left( \frac{\det (R + pHH^T)}{\det R} \right)$$ is the Shannon capacity of the channel. 
\end{theorem}
Note that in the setting of Theorem \ref{thm: SIMO}, the decoder utilizes the vector of $m$ channel outputs to estimate the source. Also, in both cases (MISO and SIMO), if the state evolution matrix $A$ is strictly stable, the source can be estimated even without communicating over the channel. In this case, the decoder can simply take the estimate $\hat S_t = 0$ and achieve $D < \infty$.

Particularizing either Theorem \ref{thm: MISO} or Theorem \ref{thm: SIMO} to scalar Gauss-Markov sources that are transmitted over a scalar AWGN channel recovers the classical result from \cite{gorbunov} that the infinite horizon error is finite if and only if $\log |a| \leq C,$ where $C = \frac{1}{2} \log (1 + \frac{p}{r})$ is the capacity of the channel \cite{tatikonda, sahaimitter, Khina}.

Kostina and Hassibi \cite[Th. 1, Th. 3, Prop. 1]{kostinahassibirate} show the necessity of \eqref{miso_res} for bounded estimation error regardless of the dimensions $k, m, n$ by deriving a lower bound to the rate-cost tradeoff of estimating and controlling an unstable linear system across an arbitrary communication channel. By taking the mean squared cost to infinity in Th. 1 of \cite{kostinahassibirate}, a finite mean squared error is achievable only if the rate exceeds the left hand side of \eqref{miso_res} and \eqref{simo_res}. Theorems \ref{thm: MISO} and \ref{thm: SIMO} establish a tight sufficiency result using a time-invariant encoder for the two special cases of MISO and SIMO AWGN channels.

\section{Proof Sketches} \label{proofs}
In this section, we establish the optimal encoder and outline the proofs for the main results.

We begin by proposing a linear encoder of the form 
\begin{equation} \label{encoding}
    \bfX_t = \Omega_t(\bfS^{t}) + \Lambda_t (\bfY^{t-1}) + \bfM_t,
\end{equation}
where $\Omega_t, \Lambda_t$ are linear mappings and $\bfM_t$ are Gaussian random variables that are independent of $(\bfS^t,\bfY^{t-1})$.

In the case where $k=n=m=1$, we recover the known fact \cite{Khina, tatikonda} that linear encoders are optimal for scalar Gauss-Markov sources and scalar AWGN channels, which follows from the solution to the causal rate-distortion function in \cite{gorbunov}.

The next result further refines the encoding structure. The main idea is that encoding the innovations process of the state is sufficient. Such encoders are often used for Gaussian channels with feedback \cite{SK2, kim2, oron}.

\begin{lemma} (Simplified encoding structure) \label{optimalencoder} \\
The optimal linear encoder can be written as
\begin{equation} \label{optencoding}
\bfX_t = \Gamma_t P_t^{-1} (\bfS_t-\hat{\bfS}_{t}),
\end{equation}
where $\hat{\bfS}_t$ is the optimal decoders' estimate, given recursively by
\begin{equation}
        \hat{\bfS}_{t} = A \hat{\bfS}_{t-1} + \mathrm K_t (\bfY_t - \E [\bfY_t | \bfY^{t-1}]),
\end{equation}
where $K_t = A\Gamma^T H^T \left( H \Gamma_t P_t^{-1} \Gamma_t^T H^T + R \right)^{-1}$, $\hat{S}_0=0$, and the error covariance $P_t = \Cov(\bfS_t - \hat \bfS_t)$.
The matrices $\{\Gamma_t\}_{t=0}^T$ must be selected to satisfy the power constraint \eqref{powerconstraint}.
\end{lemma}
The encoding matrix $\Gamma_t \in \mathbb{R}^{n \times k}$ can be designed prior to communication since it depends on the source and channel parameters, but not on their realizations.
\begin{proof}
    First, we define an encoder $\bfX_t = \Gamma_t P_t^{-1}(\bfS_t - \hat S_t) + M_t$, where $\bfM_t \sim \mathcal{N} (0, G_t)$ and call it policy $V$. Note that $P_t$ is invertible by Assumption \ref{asm: controllable}. Let policy $U$ represent a general linear encoder of the form \eqref{encoding}. Let $\Xi^{U/V}_t \triangleq (S_t - \hat S_{t}, X_t)^T$ be a Gaussian random vector induced by either policy $U$ or $V$. We show via induction that for any $\{ \Xi^U_t \}_{t<T}$, we can construct $\{ \Xi^V_t \}_{t<T}$ to have the same distribution. Specifically, we construct
    \begin{dmath}
    \Gamma_t = \E_U\left[\bfX_t(\bfS_t - \hat \bfS_t)^T\right], \label{gt}
    \end{dmath}
    and
    \begin{dmath}
    G_t = \E_U[\bfX_t\bfX_t^T] - \E_U\left[\bfX_t(\bfS_t - \hat \bfS_t)^T\right] P_t^{-1} \E_U\left[(\bfS_t - \hat \bfS_t)\bfX_t^T\right], \label{mt}
    \end{dmath}
    in terms of policy $U$. This shows that the optimal prediction error, $P_t$, from a general encoder can be achieved with a simplified policy $V$ at every time. Finally, note that in the objective, the addition of the mean term $M_t$ is essentially noise. It cannot improve the estimation error and can only increase the power. Thus, we set $M_t = 0, \forall t$, resulting in the final encoder.
    
    The decoder estimate is given by the Kalman filter.
\end{proof}
Intuitively, at every time, $t$, the decoder needs knowledge of the most recent source state $S_t$. The encoder can communicate what is currently unknown to the decoder with minimal power by transmitting the innovation $S_t - \hat S_{t}$, where decoder prediction $\hat S_{t}$ can be computed using channel feedback.

The linear encoder-decoder pair we propose is optimal among encoders of the form \eqref{encoding}; it achieves the minimum finite horizon average estimation error: $\frac{1}{T} \sum_{t=0}^{T-1} D_t$. 

We apply a time-invariant encoding structure: 
\begin{equation} \label{inv_encoding}
    \bfX_t = \Gamma P_t^{-1} (\bfS_t - \hat \bfS_{t}).
\end{equation}
We will proceed to show that such a time-invariant linear encoder is sufficient to achieve the minimum capacity required for finite estimation error.

Using this encoder, the estimation error at the infinite horizon is given by the Riccati recursions in Lemma \ref{lemma: riccati}.

\begin{lemma}  \label{lemma: riccati}(Induced Riccati Recursions and Convergent Behavior)
The prediction error covariance, $P_t$ defined in \eqref{cov},
of the optimal code introduced in Lemma \ref{optimalencoder} evolves according to the Riccati recursion
\begin{dmath} \label{riccati_recursion}
    P_{t+1} = AP_tA^T  + Q - A\Gamma^T H^T (R+H \Gamma P_t^{-1} \Gamma^T H^T )^{-1}H\Gamma A^T 
\end{dmath}
The solution $P_t$ to \eqref{riccati_recursion} converges to the trace-maximizing or stabilizing solution $P$ of the discrete algebraic Riccati equation \cite[Sec. E.4]{linearestimation}:
\begin{dmath} \label{DARE}
P = APA^T  + Q - A\Gamma^T H^T (R+H \Gamma P^{-1} \Gamma^T H^T )^{-1}H\Gamma A^T. 
\end{dmath}
Where $\Gamma$ satisfies the power constraint 
\begin{equation} \label{base_power}
    \Tr (\Gamma P^{-1} \Gamma^T) \leq p.
\end{equation}

\begin{proof}
    The time-invariant coding scheme \eqref{inv_encoding} induces the system
    \begin{align}
        \bfS_{t+1} &= A\bfS_t + W_t \\
        \bfY_t &= H \Gamma P_t^{-1}(\bfS_t - \hat \bfS_{t}) + \bfZ_t
    \end{align}
    From classical linear estimation theory, the error covariance $P_t$ follows the Riccati recursions \eqref{riccati_recursion}. The convergence of $P_t$ to $P$ follows from \cite[Lemma E.2.1]{linearestimation}.
    
    Because of the convergence of $P_t$, the infinite horizon power constraint \eqref{base_power} is equivalent to the limit of \eqref{powerconstraint} as we take $T \to \infty$.
\end{proof}
\end{lemma}

It follows from Lemma \ref{lemma: riccati} that the infinite horizon estimation error is finite if and only if there exists a $\Gamma$ such that a positive semidefinite $P$ satisfies both \eqref{DARE} and \eqref{base_power}.

So far, we have made no assumptions about the dimensions of the channel and source. Next, we derive the main results, which require dimensionality assumptions.

\subsection{Proof of Theorem \ref{thm: MISO}}
\subsubsection{Preliminaries}
Here, $H \in \mathbb{R}^{1 \times n}, A \in \mathbb{R}^{k \times k}$, and $||H|| = 1$. We will show that $D < \infty$ if and only if
\begin{equation}
    \frac{1}{2}\log(1+\frac{p}{r}) > \sum_{i=1}^k \log \left( \max (1, |\lambda_i|) \right)
\label{bounded_cond}
\end{equation}
The channel input $X_t$ can be decomposed into orthogonal components in the row and null space of $H$. The component in the null space is nulled by the channel. Thus, to preserve power, we should only optimize over channel inputs of the form
\begin{equation}
    \bfX_t = H^T \Gamma P^{-1}(\bfS_t - \hat \bfS_t),
\end{equation}
where Assumption \ref{asm: controllable} guarantees $P \succ 0$. Note that the power constraint is invariant due to the assumption on the norm of $H$. Here, $\Gamma \in \mathbb{R}^{1 \times k}$. From Lemma \ref{lemma: riccati}, this has a corresponding DARE
\begin{equation} \label{miso_dare}
    P = APA^T + Q -  A \Gamma^T  (r + \Gamma P^{-1} \Gamma^T)^{-1} \Gamma A^T
\end{equation}
with power constraint
\begin{equation} \label{miso_power}
    \Gamma P^{-1} \Gamma^T \leq p.
\end{equation}

Since the optimal encoder utilizes all available power, we reduce the DARE \eqref{miso_dare} to a Lyapunov equation by applying the power constraint with equality. The error covariance is given by the solution to
\begin{equation} \label{lyap_p}
    P = APA^T + Q - \frac{A \Gamma^T \Gamma A^T}{r + p},
\end{equation}
where $r$ is the channel noise variance and $p$ is the transmit power. The power constraint \eqref{miso_power} is equivalent to
\begin{equation} \label{proof: p_cons}
    \left[ \begin{array}{cc} P & \Gamma^T \\ \Gamma & p \end{array} \right] \succeq 0,
\end{equation}
by the Schur complement lemma \cite[Sec. C.4]{boydconvex}, which is, in turn, equivalent to
\begin{equation} \label{proof: J_cons}
J \triangleq P -\frac{\Gamma^T\Gamma}{p} \succeq 0,
\end{equation}
\begin{equation}
    P = J+\frac{\Gamma^T\Gamma}{p}.
\end{equation}
Plugging this into \eqref{lyap_p}, we get
\begin{equation} \label{proof: J_simp}
    J+\frac{\Gamma^T\Gamma}{p} = A\left(J+\frac{\Gamma^T\Gamma}{p}\right)A^T+Q-\frac{A\Gamma^T\Gamma A^T}{r+p},
\end{equation}
which, upon simplification, yields the following Lyapunov equation for $J$:
\begin{equation}
J = AJA^T+Q-\frac{\Gamma^T\Gamma}{p}+\frac{A\Gamma^T\Gamma A^T}{p(1+p/r)}. 
\label{lyap_r}
\end{equation}
There always exists a solution to $J$ due to Assumption \eqref{asm: eigs}. Therefore $P$ is finite if and only if there exists a $\Gamma$ such that $J \succeq 0$. We proceed to show the latter.

From Assumption \ref{asm: diag},
\[ A = \left[ \begin{array}{cc} A_s & 0 \\ 0 & A_u \end{array} \right], \]
where $A_s$ is stable and $A_u$ is unstable and diagonal.

\subsubsection{Proof of Sufficiency}
We will show that if (\ref{bounded_cond}) holds, then there exists a $\Gamma$ such that the solution, $J$, to \eqref{lyap_r} is PSD. To this end, let us assume that $\Gamma$ takes the form
\[ \Gamma =\begin{bmatrix} 0  & \Gamma_u \end{bmatrix}, \] so that we send no information about the stable components of the source state. In this case, (\ref{lyap_r}) takes the form
\begin{dmath}
\left[ \begin{array}{cc} J_{ss} & J_{su} \\ J_{us} & J_{uu} \end{array} \right] = \left[ \begin{array}{cc} A_s & 0 \\ 0 & A_u \end{array} \right]
\left[ \begin{array}{cc} J_{ss} & J_{su} \\ J_{us} & J_{uu} \end{array} \right] \left[ \begin{array}{cc} A_s^T & 0 \\ 0 & A_u^T \end{array} \right]
+ \left[ \begin{array}{cc} Q_{ss} & Q_{su} \\ Q_{us} & Q_{uu} \end{array} \right] +
\left[ \begin{array}{cc} 0 & 0 \\ 0 & -\frac{\Gamma_u^T\Gamma_u}{p} + \frac{A_u\Gamma_u^T\Gamma_u A_u^T}{p(1+p/r)} \end{array} \right] .
\label{block_r}
\end{dmath}

Let us assume that $(A_s,Q_s)$ is controllable, which follows from the controllability of $(A,Q)$, i.e. Assumption \ref{asm: controllable}. Then, since $J_{ss} = A_sJ_{ss}A_s^T+Q_s$ and $A_s$ is stable, $Q_s \succ 0$, it follows that $J_{ss} \succ 0$ by the Lyapunov stability theorem \cite[Lemma D.1.2]{linearestimation}. Thus, there exists a $\Gamma_u$ such that $J \succeq 0$, if and only if there exists a $\Gamma_u$ such that 
\begin{equation}
J_{uu}-J_{us}J^{-1}_{ss}J_{su} \succeq 0.
\label{schur_comp}
\end{equation}
To show this, let us focus on the equation for $J_{uu}$:
\begin{equation}
    J_{uu} = A_uJ_{uu}A_u^T+Q_{uu} -\frac{\Gamma_u^T \Gamma_u}{p} + \frac{A_u\Gamma_u^T \Gamma_u A_u^T}{p(1+p/r)}
\end{equation}
which we rearrange as
\begin{dmath}
J_{uu} = A_u^{-1}J_{uu}A_u^{-T}-A_u^{-1}Q_{uu}A_u^{-T} + \frac{A_u^{-1}\Gamma_u^T \Gamma_u A_u^{-T}}{p}  - \frac{\Gamma_u^T \Gamma_u}{p(1+p/r)}  .
\label{lyap_ru}
\end{dmath}
Note that $A_u^{-1}$ is now stable. 

Defining $D_u \triangleq \mbox{diag}(\Gamma_u)$, i.e., the diagonal matrix whose components are the elements of the vector $\Gamma_u^T$, we may now write
\[ A_u^{-1}\Gamma_u^T = D_u a ~~~\mbox{and}~~~\Gamma_u^T = D_u1, \]
where $a$ is the vector of the diagonal elements of $A_u^{-1}$ and $1$ is the all-one vector. Without Assumption \ref{asm: diag}, $A_u$ is in block Jordan form, which results in a non-diagonal $D_u$ and a more complex expression for $\Gamma$. In the diagonal case, (\ref{lyap_ru}) becomes
\begin{dmath}
J_{uu} = A_u^{-1}J_{uu}A_u^{-T}-A_u^{-1}Q_{uu}A_u^{-T} + D_u\left(\frac{aa^T}{p}-\frac{11^T}{p(1+p/r)} \right)D_u .
\label{lyap_ru2}
\end{dmath}

Because the above Lyapunov equation for $J_{uu}$ is linear, its solution can be written as $J_{uu} = {\hat J}_{uu}+{\tilde J}_{uu}$ where
\begin{equation}
{\hat J}_{uu} = A_u^{-1}{\hat J}_{uu}A_u^{-T}-A_u^{-1}Q_{uu}A_u^{-T},
\label{lyap_hat}
\end{equation}
and 
\begin{equation}
{\tilde J}_{uu} = A_u^{-1}{\tilde J}_{uu}A_u^{-T} + D_u\left(\frac{aa^T}{p}-\frac{11^T}{p(1+p/r)} \right)D_u .
\label{lyap_tilder}
\end{equation}
Note that, because $A_u^{-1}$ is stable and $Q_{uu} \succeq 0$, we have by \eqref{lyap_hat} and the Lyapunov stability theorem that ${\hat J}_{uu} \preceq 0.$ Thus for (\ref{schur_comp}) to hold, ${\tilde J}_{uu}$ must be sufficiently PSD. 

Let $M$ be the solution to the Lyapunov equation
\begin{equation}
M = A_u^{-1}MA_u^{-T}+11^T.
\label{lyap_m}
\end{equation}
Let $(A_u, 1)$ be controllable. Then by the Lyapunov stability theorem, $M \succ 0$. We now claim that 
\begin{equation}
{\tilde J}_{uu} = D_u\left(\frac{M}{r+p}-\frac{11^T}{p}\right)D_u. 
\label{tilder}
\end{equation}
This can be verified by plugging (\ref{tilder}) into (\ref{lyap_tilder}).
It follows that ${\tilde J}_{uu} \succ 0$ if and only if $\frac{M}{r+p}-\frac{11^T}{p}>0$. But the latter is equivalent to
\begin{equation}
    \left[\begin{array}{cc} M & 1 \\ 1^T & \frac{p}{r+p} \end{array} \right] \succ 0,
\end{equation}
or
\begin{equation}
    \frac{p}{r+p}>1^TM^{-1}1.
\label{power_done}
\end{equation}

Assume $M$ satisfies the Lyapunov equation (\ref{lyap_m}). Then
\begin{equation}
    1^TM^{-1}1 = 1-\left|\mbox{det}(A_u)\right|^{-2}.
    \label{M_det}
\end{equation}
This follows since $A_u^{-1}MA_u^{-T} = M-11^T$ from \eqref{lyap_m}. On the one hand
\begin{equation}
    \mbox{det}A_u^{-1}MA_u^{-T} = \left|\mbox{det}A_u\right|^{-2}\mbox{det}{M}
\end{equation}
and on the other
\begin{dmath}
    \mbox{det}(M-11^T) = \mbox{det}(I-11^TM^{-1}) \ \mbox{det} M  = (1-1^TM^{-1}1) \ \mbox{det}M,
\end{dmath}
which yields the desired result.

Then, \eqref{power_done} and \eqref{M_det} imply that $J_{uu} \succ 0$ if and only if
\begin{equation}
    \frac{p}{r+p}>1-\left|\mbox{det}A_u\right|^{-2},
\end{equation}
or equivalently
\begin{equation} \label{cap_cond}
    1+\frac{p}{r} > \left|\mbox{det}A_u\right|^2,
\end{equation}
which is the capacity condition we are seeking.

Note that when this capacity condition holds, $\frac{M}{r+p}-\frac{11^T}{p} \succ 0$ and therefore ${\tilde J}_{uu} \succ 0$ in \eqref{tilder}. Moreover, since ${\tilde J}_{uu}$ is scaled on both sides by $D_u$ (whose entries are the elements of $\Gamma_u$) we can choose $\Gamma$ to make it arbitrarily positive definite, which means we can always satisfy (\ref{schur_comp}).

This establishes sufficiency.

\subsubsection{Proof of Necessity}
For necessity, note that if we now take an arbitrary 
\[ \Gamma =\begin{bmatrix} \Gamma_s  & \Gamma_u \end{bmatrix}, \]
i.e. we do not set $\Gamma_s$ to zero, then the Lyapunov equation for $J_{uu}$ does not change. This means that it continues to hold in \eqref{lyap_r} that $J_{uu} = {\hat J}_{uu}-{\tilde J}_{uu}$, where ${\hat J}_{uu}$ is negative definite. When the capacity condition \eqref{cap_cond} is violated then $\frac{M}{r+p}-\frac{11^T}{p}$, where $M$ is unaffected by $\Gamma$, has at least one negative eigenvalue which implies that, no matter what $D_u$ is, ${\tilde J}_{uu}$ will have at least one negative eigenvalue. Therefore the same must be true of $J_{uu}$ and so $J$ cannot be positive definite, meaning that $P$ will not be bounded.

This establishes necessity.

\subsection{Proof of Theorem \ref{thm: SIMO}}
Here, $H \in \mathbb{R}^m$ is a vector and $A \in \mathbb{R}^{k \times k}$. From Lemma \ref{lemma: riccati}, this has a corresponding DARE
\begin{equation}
    P = APA^T + Q - A \Gamma^T H^T(r + H \Gamma P^{-1} \Gamma^T H^T)^{-1}H \Gamma A^T
\end{equation}
with power constraint
\begin{equation}
    \Gamma P^{-1} \Gamma^T \leq p.
\end{equation}

We apply the same reasoning as the proof for Theorem \ref{thm: MISO} to bound the channel input power, resulting in the Lyapunov equation:

\begin{equation}
    P = APA^T + Q - H^T(R+HH^Tp)^{-1}H A\Gamma^T\Gamma A^T.
\end{equation}

Applying the power constraint using the method of \eqref{proof: p_cons} and \eqref{proof: J_cons}, we obtain
\begin{equation} \label{proof: J_fin2}
    J = AJA^T + Q - \frac{\Gamma^T \Gamma}{p} + \frac{1}{p(1+H^TR^{-1}H p)} A\Gamma^T \Gamma A^T.
\end{equation}
\eqref{proof: J_fin2} parallels \eqref{lyap_r} for the MISO case. Applying the proof steps that follow \eqref{lyap_r}, we conclude that 
$$ 1 + H^T R^{-1} H p > (\det A_u)^2$$
is a necessary and sufficient condition for finite estimation error at the infinite horizion.
Rearranging terms, we arrive at the desired capacity condition of Theorem \eqref{thm: SIMO}.

\section{Conclusion and Future Work}
In this paper, we analyzed a zero-delay communication system composed of a potentially vector-valued Gauss-Markov source and AWGN channel. We derived necessary and sufficient conditions for the mean square error to be finite at the infinite horizon for SIMO and MISO AWGN channels in Theorem \ref{thm: MISO} and Theorem \ref{thm: SIMO} respectively. This revealed a strong connection between the instability of a linear system, quantified by its unstable eigenvalues, and the Shannon capacity of the communication channel.

Generalizing these results to general MIMO channels is of significant interest and is the subject of ongoing research. The current paper takes a significant step towards this characterization as the linear encoder and decoder in Lemma \ref{optimalencoder} are optimal for MIMO channels even in some cases where the dimensions are unmatched \cite{ZAIDI201632}. Furthermore, the finite-horizon problem with a time-varying encoder is fundamentally connected to the problems of automatic control and anytime coding. In a control setting, the minimum mean square estimate can be used to control the unstable system. In an anytime communication setting, the source can be modeled as an information source containing a stream of bits that arrives gradually. An accurate estimate of the source state will allow us to determine and decode the transmitted bits in a streaming fashion.

\bibliographystyle{ieeetr}
\bibliography{ref} % Entries are in the refs.bib file

\begin{thebibliography}{10}

\bibitem{shannon}
C.~E. Shannon, ``A mathematical theory of communication,'' {\em The Bell System
  Technical Journal}, vol.~27, pp.~379--423, Oct. 1948.

\bibitem{hamming}
R.~W. Hamming, ``Error detecting and error correcting codes,'' {\em The Bell
  System Technical Journal}, vol.~29, pp.~147--160, Apr. 1950.

\bibitem{reed}
I.~S. Reed and G.~Solomon, ``Polynomial codes over certain finite fields,''
  {\em Journal of the Society for Industrial and Applied Mathematics}, vol.~8,
  pp.~300--304, June 1960.

\bibitem{reedmuller}
I.~Reed, ``A class of multiple-error-correcting codes and the decoding
  scheme,'' {\em Transactions of the IRE Professional Group on Information
  Theory}, vol.~4, pp.~38--49, Sep. 1954.

\bibitem{polar}
E.~Arikan, ``Channel polarization: A method for constructing capacity-achieving
  codes for symmetric binary-input memoryless channels,'' {\em IEEE
  Transactions on Information Theory}, vol.~55, pp.~3051--3073, June 2009.

\bibitem{ldpc}
R.~Gallager, ``Low-density parity-check codes,'' {\em IRE Transactions on
  Information Theory}, vol.~8, pp.~21--28, Jan. 1962.

\bibitem{horstein}
M.~Horstein, ``Sequential transmission using noiseless feedback,'' {\em IEEE
  Transactions on Information Theory}, vol.~9, pp.~136--143, July 1963.

\bibitem{SK2}
J.~Schalkwijk and T.~Kailath, ``A coding scheme for additive noise channels
  with feedback--i: No bandwidth constraint,'' {\em IEEE Transactions on
  Information Theory}, vol.~12, pp.~172--182, Apr. 1966.

\bibitem{NaghshvarSED}
M.~Naghshvar, M.~Wigger, and T.~Javidi, ``Optimal reliability over a class of
  binary-input channels with feedback,'' in {\em 2012 IEEE Information Theory
  Workshop}, pp.~391--395, 2012.

\bibitem{shannonfeedback}
C.~E. Shannon, ``The zero error capacity of a noisy channel,'' {\em IRE
  Transactions on Information Theory}, vol.~2, pp.~8--19, Sep. 1956.

\bibitem{nian2}
N.~Guo and V.~Kostina, ``Reliability function for streaming over a dmc with
  feedback,'' {\em IEEE Transactions on Information Theory}, vol.~69,
  pp.~2165--2192, Nov. 2023.

\bibitem{nian}
N.~Guo and V.~Kostina, ``Instantaneous {SED} coding over a {DMC},'' in {\em
  2021 IEEE International Symposium on Information Theory (ISIT)},
  pp.~148--153, 2021.

\bibitem{sahaimitter}
A.~Sahai and S.~Mitter, ``The necessity and sufficiency of anytime capacity for
  stabilization of a linear system over a noisy communication link—part i:
  Scalar systems,'' {\em IEEE Transactions on Information Theory}, vol.~52,
  pp.~3369--3395, July 2006.

\bibitem{Khina}
A.~Khina, E.~R. Gårding, G.~M. Pettersson, V.~Kostina, and B.~Hassibi,
  ``Control over {G}aussian channels with and without source–channel
  separation,'' {\em IEEE Transactions on Automatic Control}, vol.~64,
  pp.~3690--3705, Apr. 2019.

\bibitem{elia}
N.~Elia, ``When {B}ode meets {S}hannon: control-oriented feedback communication
  schemes,'' {\em IEEE Transactions on Automatic Control}, vol.~49,
  pp.~1477--1488, Sep. 2004.

\bibitem{skoglund}
P.~A. Floor, A.~N. Kim, N.~Wernersson, T.~A. Ramstad, M.~Skoglund, and
  I.~Balasingham, ``Zero-delay joint source-channel coding for a bivariate
  {G}aussian on a {G}aussian {MAC},'' {\em IEEE Transactions on
  Communications}, vol.~60, pp.~3091--3102, July 2012.

\bibitem{schulman}
L.~Schulman, ``Coding for interactive communication,'' in {\em Proceedings of
  1995 IEEE International Symposium on Information Theory}, pp.~452--, Nov.
  1995.

\bibitem{sukhanytime}
R.~T. Sukhavasi and B.~Hassibi, ``Linear time-invariant anytime codes for
  control over noisy channels,'' {\em IEEE Transactions on Automatic Control},
  vol.~61, pp.~3826--3841, Feb. 2016.

\bibitem{gorbunov}
A.~Gorbunov and M.~S. Pinsker, ``Prognostic epsilon entropy of a {G}aussian
  message and a {G}aussian source,'' {\em Problemy Peredachi Informatsii},
  vol.~10, pp.~5--25, Aug. 1974.

\bibitem{kostinahassibirate}
V.~Kostina and B.~Hassibi, ``Rate-cost tradeoffs in control,'' {\em IEEE
  Transactions on Automatic Control}, vol.~64, pp.~4525--4540, Apr. 2019.

\bibitem{Tanaka}
T.~Tanaka, K.-K.~K. Kim, P.~A. Parrilo, and S.~K. Mitter, ``Semidefinite
  programming approach to {G}aussian sequential rate-distortion trade-offs,''
  {\em {IEEE} Transactions on Automatic Control}, vol.~62, pp.~1896--1910, Apr.
  2017.

\bibitem{tatikonda}
S.~Tatikonda, A.~Sahai, and S.~Mitter, ``Stochastic linear control over a
  communication channel,'' {\em IEEE Transactions on Automatic Control},
  vol.~49, pp.~1549--1561, Sep. 2004.

\bibitem{gastpar}
M.~Gastpar, B.~Rimoldi, and M.~Vetterli, ``To code, or not to code: lossy
  source-channel communication revisited,'' {\em IEEE Transactions on
  Information Theory}, vol.~49, pp.~1147--1158, May 2003.

\bibitem{nair}
G.~N. Nair, F.~Fagnani, S.~Zampieri, and R.~J. Evans, ``Feedback control under
  data rate constraints: An overview,'' {\em Proceedings of the IEEE}, vol.~95,
  pp.~108--137, Jan. 2007.

\bibitem{nair2}
G.~N. Nair and R.~J. Evans, ``Stabilizability of stochastic linear systems with
  finite feedback data rates,'' {\em SIAM Journal on Control and Optimization},
  vol.~43, pp.~413--436, July 2004.

\bibitem{borkar}
V.~S. Borkar, ``Control of markov chains with long-run average cost
  criterion,'' in {\em Stochastic Differential Systems, Stochastic Control
  Theory and Applications} (W.~Fleming and P.-L. Lions, eds.), (New York, NY),
  pp.~57--77, Springer New York, 1988.

\bibitem{basar}
O.~C. Imer, S.~Yüksel, and T.~Başar, ``Optimal control of {LTI} systems over
  unreliable communication links,'' {\em Automatica}, vol.~42, pp.~1429--1439,
  Sep. 2006.

\bibitem{franceschetti}
L.~Schenato, B.~Sinopoli, M.~Franceschetti, K.~Poolla, and S.~S. Sastry,
  ``Foundations of control and estimation over lossy networks,'' {\em
  Proceedings of the IEEE}, vol.~95, pp.~163--187, Jan. 2007.

\bibitem{gupta}
V.~Gupta, D.~Spanos, B.~Hassibi, and R.~Murray, ``On {LQG} control across a
  stochastic packet-dropping link,'' in {\em Proceedings of the 2005, American
  Control Conference, 2005.}, pp.~360--365 vol. 1, 2005.

\bibitem{zaidi2013stabilization}
A.~A. Zaidi, T.~J. Oechtering, S.~Yuksel, and M.~Skoglund, ``Stabilization of
  linear systems over {G}aussian networks,'' 2013.

\bibitem{linearestimation}
T.~Kailath, A.~Sayed, and B.~Hassibi, {\em Linear Estimation}.
\newblock Prentice-Hall information and system sciences series, Prentice Hall,
  2000.

\bibitem{kim2}
Y.-H. Kim, ``Feedback capacity of stationary gaussian channels,'' {\em IEEE
  Transactions on Information Theory}, vol.~56, no.~1, pp.~57--85, 2010.

\bibitem{oron}
O.~Sabag, V.~Kostina, and B.~Hassibi, ``Feedback capacity of {MIMO} {G}aussian
  channels,'' {\em IEEE Transactions on Information Theory}, vol.~69, no.~10,
  pp.~6121--6136, 2023.

\bibitem{boydconvex}
S.~Boyd and L.~Vandenberghe, {\em Convex optimization}.
\newblock Cambridge university press, 2004.

\bibitem{ZAIDI201632}
A.~A. Zaidi, S.~Yüksel, T.~J. Oechtering, and M.~Skoglund, ``On the tightness
  of linear policies for stabilization of linear systems over {G}aussian
  networks,'' {\em Systems and Control Letters}, vol.~88, pp.~32--38, 2016.

\end{thebibliography}

\end{document}